\documentclass{article}

\usepackage{graphicx}

\def\abstract#1{\vskip 7mm
        \begin{center}{\large Abstract}\par \smallskip
                \begin{minipage}[c]{12cm}
                        \small #1
                \end{minipage}
        \end{center}
}
\def\title#1{\begin{center}{\Large\bf #1}\end{center}}
\def\author#1{\vskip 5mm \begin{center}{#1}\end{center}}
\def\address#1{\begin{center}{\it #1}\end{center}}
\makeatletter
\def\vereq#1#2{\lower3pt\vbox{\baselineskip1.5pt \lineskip1.5pt
\ialign{$\m@th#1\hfill##\hfil$\crcr#2\crcr\sim\crcr}}}
\makeatother

\def\beq{\begin{equation}}
\def\eeq{\end{equation}}
\def\ber{\begin{eqnarray}}
\def\eer{\end{eqnarray}}
\def \lleq {\lower0.9ex\hbox{ $\buildrel < \over \sim$} ~}
\def \ggeq {\lower0.9ex\hbox{ $\buildrel > \over \sim$} ~}

\def\ie{{\it i.e.~}}
\def\statei{\lbrace r,s \rbrace}
\def\statej {\lbrace r,q \rbrace}

\def\atridot{\stackrel{...}{a}}

\def\omx {\Omega_X}
\def\om {\Omega_m}

\def\half{{1\over 2}}

\def\lsim{\
\lower-1.5pt\vbox{\hbox{\rlap{$<$}\lower5.3pt\vbox{\hbox{$\sim$}}}}\ }
\def\gsim{\
\lower-1.5pt\vbox{\hbox{\rlap{$>$}\lower5.3pt\vbox{\hbox{$\sim$}}}}\ }

\def\apj{{Astroph.\@ J.\ }}

\def\aj{{Astron.\@ J.\ }}

\def\prd{{Phys.\@ Rev.\@ D\ }}
\def\pd{{Phys.\@ Rev.\@ D\ }}

\def\plb {{Phys.\@ Lett.\@ B\ }}
\def \jetpl {JETP Lett.\ }
\def\etal{{\it et al.}}

\begin{document}

\title{Cosmological Surprises from Braneworld models of Dark Energy}
\author{%
  Varun Sahni\footnote{E-mail:varun@iucaa.ernet.in}
}
\address{%
Inter-University Centre for Astronomy and Astrophysics,
Pune 411 007, India
}

\abstract{Properties of Braneworld models of dark energy are reviewed.
Braneworld models admit the following interesting possibilities: (i) The
effective equation of state can be $w \leq -1$ as well as $w \geq -1$. In the
former case the expansion of the universe is well behaved at all times and the
universe does not run into a future `Big Rip' singularity which is usually
encountered by Phantom models. (ii) For a class of Braneworld models the
acceleration of the universe can be a {\em transient\/} phenomenon. In this
case the current acceleration of the universe is sandwiched between two matter
dominated epochs. Such a braneworld does not have a horizon in contrast to LCDM
and most Quintessence models. (iii) For a specific set of parameter values the
universe can either originate from, or end its existence in a {\em Quiescent\/}
singularity, at which the density, pressure and Hubble parameter {\em remain
finite\/}, while the deceleration parameter and all invariants of the Riemann
tensor diverge to infinity within a finite interval of cosmic time. (iv)
Braneworld models of dark energy can {\em loiter\/} at high redshifts: $6 \lleq
z \lleq 40$. The Hubble parameter {\em decreases\/} during the loitering epoch
relative to its value in LCDM. As a result the age of the universe at loitering
dramatically increases and this is expected to boost the formation of high
redshift gravitationally bound systems such as $10^9 M_\odot$ black holes at $z
\sim 6$ and lower-mass black holes and/or Population III stars at $z > 10$,
whose existence could be problematic within the LCDM scenario. (v) Braneworld
models with a time-like extra dimension bounce at early times thereby avoiding
the initial `Big Bang singularity'.
(vi) Both Inflation and Dark Energy can be successfully unified within
a single scheme (Quintessential Inflation).
}

\section{Introduction}

One of the most remarkable discoveries of the past decade is that the universe
is accelerating. Evidence for an accelerating universe comes from observations
of high redshift type Ia supernovae treated as standardized candles \cite{sn}
and, more indirectly, by observations of the cosmic microwave background and
galaxy clustering \cite{wmap,tegmark03}. Perhaps the simplest explanation for
acceleration is the presence of vacuum energy exhibiting itself as a small
cosmological constant with equation of state $p = -\rho$ = constant. However,
its un-evolving nature implies that the cosmological constant must be set to an
extremely small value in order to dominate the expansion dynamics of the
universe at precisely the present epoch and this gives rise (according to one's
perspective) either to an initial `fine-tuning' problem or to a `cosmic
coincidence' problem. As a result several radically different alternative
methods of generating `dark energy' at a sufficiently late cosmological epoch
have been suggested (see \cite{ss00,carroll01,pr02,paddy03,sahni04} for reviews
on this subject). In this talk I will focus on one such approach which rests on
the notion that space-time is higher-dimensional, and that our observable
universe is a (3+1)-dimensional `brane' which is embedded in a
(4+1)-dimensional `bulk' space-time. As we shall see, higher-dimensional
braneworld models allow the expansion dynamics to be radically different from
that predicted by conventional Einstein gravity in 3+1 dimensions. Some
cosmological `surpises' which spring from Braneworld models include:

\begin{itemize}

\item Both early and late time acceleration can be successfully unified within
a single scheme (Quintessential Inflation) in which the very same scalar
field which drives Inflation at early times becomes Quintessence at
late times.

\item The (effective) equation of state of dark energy in the braneworld
scenario can be `phantom-like' ($w < - 1$) or `Quintessence-like' ($w > - 1$).
(These two possibilities are essentially related to the two
distinct ways in which the brane can be embedded in the bulk.)

\item The acceleration of the universe can be a {\em transient\/} phenomenon:
braneworld models accelerate during the present epoch but return to
matter-dominated expansion at late times.

\item A class of braneworld models encounter a {\em Quiescent future
singularity}, at which ${\dot a} \to $ constant,
but ${\ddot a} \to -\infty $. The surprising feature of this singularity
is that while the Hubble parameter, density and pressure remain finite, the
deceleration parameter and all curvature invariants {\em diverge\/} as the
singularity is approached.

\item A spatially flat Braneworld can {\em mimick\/} a closed universe and {\em
loiter\/} at large redshifts.

\item A braneworld embedded in a five-dimensional space in which the extra
(bulk) dimension is time-like can {\em bounce at early times}, thereby
generically avoiding the big bang singularity. Cyclic models of the universe
with successive expansion-contraction cycles can be constructed based on such a
bouncing braneworld.

\end{itemize}

Let us now dwell a little on each of these cosmological properties (`surprises').

\section{Quintessential Inflation on the Brane}

An intriguing issue in cosmology is that the universe appears to accelerate
twice: once at the very beginning during Inflation and again about
10 billion years later, during the present epoch. Although most theoretical
models assume that there is no real connection between the two epochs and that
Inflation and Dark energy are distinct physical entities, it might well be that
the two phenomena are in fact related, and that the same scalar field which
initially drives inflation, later, when its density has been considerably
reduced, plays the role of Quintessence. This notion of Quintessential
Inflation was first explored in the context of the Einstein gravity in 3+1
dimensions by Peebles and Vilenkin \cite{pv99}. The possibility that Braneworld
models could provide a more efficient realisation of this scenario was
suggested by Copeland, Liddle and Lidsey \cite{copeland} and subsequently
discussed in greater detail by several authors
\cite{lidsey1,sss02,majumdar,nojiri03,sami}.

The main line of reasoning behind this approach is simple.
In the Randall--Sundrum model \cite{RS} the modified  Einstein equations on the
brane contain high-energy corrections as well as the projection of the Weyl
tensor from the bulk on to the brane. The Friedmann equation on the brane in
this case becomes \cite{BDL}
\begin{equation}\label{eq:cosmolim}
H^2 + {\kappa \over a^2} = {\Lambda_{\rm b} \over 6} + {C \over a^4} +
{\left(\rho + \sigma \right)^2 \over 9 M^6} \,  .
\end{equation}
where $\kappa = 0, \pm 1$,  $\sigma$ is the brane tension,
$\Lambda_{\rm b}$ is the value of the cosmological constant in the
five-dimensional `bulk' and $C/a^4$ is the `dark radiation' term which
describes the projected five-dimensional degrees of freedom onto the brane. It
is easy to see from (\ref{eq:cosmolim}) that
\beq
H^2(a\to\infty) = \frac{\Lambda_{\rm RS}}{3}~,
\eeq
where
\beq
\Lambda_{\rm RS} = \frac{\Lambda_{\rm b}}{2} + \frac{\sigma^2}{3M^6}
\label{eq:lam_RS}
\eeq
is the effective four dimensional cosmological constant in the Randall--Sundrum
model.

\begin{figure}[ht]
\centering
\includegraphics[width=9cm]{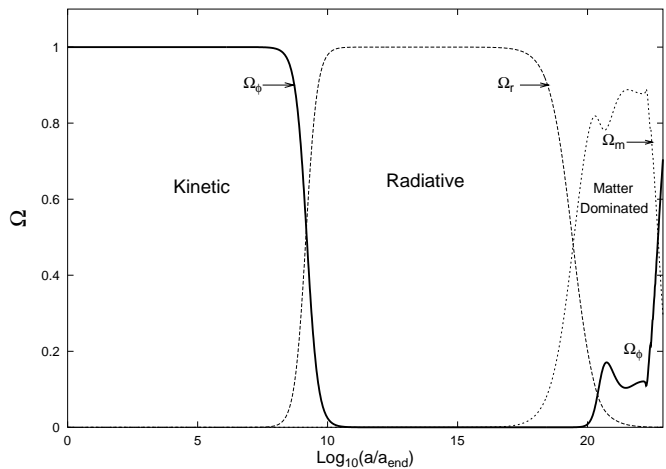}
\caption{\footnotesize The post-inflationary  density parameter $\Omega$ is
plotted for the scalar field (solid line), radiation (dashed line), and cold
dark matter (dotted line) in the quintessential-inflationary model described by
the potential ~\protect\cite{sw00} $V(\phi) = V_0[\cosh{\lambda\phi} - 1]^p$
with $p = 0.2$. Late-time oscillations of the scalar field ensure that the mean
equation of state turns negative $\langle w_\phi\rangle \simeq -2/3$, giving
rise to the current epoch of cosmic acceleration with $a(t) \propto t^2$ and
present-day values $\Omega_{0\phi} \simeq 0.7$, $\Omega_{0m} \simeq 0.3$. Note
the existence of two epochs, early and late, when the scalar field
dominates the matter density of the universe. Figure courtesy of \cite{sami}. }
\bigskip
\medskip
\label{fig:quint_inf}
\end{figure}

During Inflation the expansion of the universe is driven by a scalar field
propagating on the brane with
energy density and pressure given by
\begin{equation}
\rho_{\phi}={ \dot{\phi}^2 \over 2}+V(\phi),~~~~p_{\phi}={ \dot{\phi}^2 \over 2}-V(\phi)~,
\end{equation}
while the evolution equation for the scalar field is
\begin{equation}
\ddot{\phi}(t)+3H\dot{\phi}(t)+V_{,\phi}=0~.
\label{eq:roll}
\end{equation}
Braneworld models add an interesting new dimension to the scalar-field
dynamics. As demonstrated by (\ref{eq:cosmolim}), for $\rho \gg \sigma$ a
quadratic density term appears in the modified Friedmann equation on the brane.
This radically alters the
expansion dynamics at early epochs by speeding up the rate of expansion and
the value of the Hubble parameter, which now becomes $H \propto \rho$ instead
of the more conventional
$H \propto \sqrt{\rho}$. Since the value of $H$ affects the motion of $\phi$
through the damping term `$3H{\dot\phi}$' in (\ref{eq:roll}),
the scalar field experiences significantly greater
damping in braneworld cosmology than in conventional GR.
Consequently, inflation on the brane can be realized by very steep potentials
-- precisely those that are used to describe Quintessence. The braneworld
scenario therefore provides us with the opportunity to unify inflation  and
dark energy through the notion of quintessential inflation
\cite{pv99,maartens,copeland,lidsey1,sss02,majumdar,sami}. An example of
quintessential inflation is shown in Figure~\ref{fig:quint_inf}.

An important property of Quintessential Inflation on the brane is that
once the inflationary regime is over and braneworld corrections cease to play an
important role, the extreme steepness of the potential causes the scalar field
to plunge down its potential resulting in a `kinetic regime' prior to reheating
during which $p_\phi \simeq \rho_\phi$. It is well known that the spectrum of
relic gravity waves created quantum mechanicaly during Inflation is sensitive
to and bears an imprint of the post-inflationary equation of state \cite{star79,sahni90}.
The effect of the kinetic regime is to create a `blue' gravity wave spectrum
on short wavelength scales which is an important observational signature of
Quintessential Inflation on the brane \cite{sss02,sami}; see also \cite{giovan98}.

\section{Braneworld models of Dark Energy with $w \leq -1$ and $w \geq -1$}

A radically different way of making the universe accelerate
was suggested in \cite{DGP,ss02}.
The braneworld model which I now discuss presents a successful synthesis of
the higher-dimensional ansatzes proposed by Randall and Sundrum \cite{RS} and
Dvali, Gabadadze, and Porrati \cite{DGP}, and is described by the action
\cite{CHD}
\beq \label{action}
S = M^3  \left[ \int_{\rm bulk} \left({\mathcal R} - 2 \Lambda_{\rm b}
\right) - 2 \int_{\rm brane} K  \right] + \int_{\rm brane} \left( m^2 R -
2 \sigma \right) + \int_{\rm brane} L (h_{ab}, \phi) \, .
\eeq
Here, ${\mathcal R}$ is the scalar curvature of the five-dimensional metric
$g_{ab}$ in the bulk, and $R$ is the scalar curvature of the induced metric
$h_{ab} = g_{ab} - n_a n_b$ on the brane, where $n^a$ is the vector field of
the inner unit normal to the brane. The quantity $K = K_{ab} h^{ab}$ is the
trace of the symmetric tensor of extrinsic curvature $K_{ab} = h^c{}_a \nabla_c
n_b$ of the brane, and $L (h_{ab}, \phi)$ denotes the Lagrangian density of the
four-dimensional matter fields $\phi$ whose dynamics is restricted to the brane
(the notation and conventions are those of \cite{Wald}). Integrations over the
bulk and brane are taken with the natural volume elements $\sqrt{- g}\, d^5 x$
and $\sqrt{- h}\, d^4 x$, respectively. The constants $M$ and $m$ denote,
respectively, the five-dimensional and four-dimensional Planck masses,
$\Lambda_{\rm b}$ is the five-dimensional (bulk) cosmological constant, and
$\sigma$ is the brane tension.

An important difference between the action (\ref{action}) and that describing
the Randall--Sundrum cosmology is the presence of the term $m^2\int
R d^4x$ in (\ref{action}). This term can be thought of as
resulting from the backreaction of quantum fluctuations of matter fields
residing on the brane, and this mechanism of generating the
gravitational part of the action was proposed by Sakharov in a seminal paper
in 1967 \cite{Sakharov}. The effect of including such a term in the braneworld
action was explored in \cite{CHD,DGP,ss02}.

Action (\ref{action}) leads to the following expression for the Hubble
parameter on the brane for a {\em spatially flat\/} universe \cite{ss02}:
\begin{equation}\label{hubble}
H^2 (a)  = {A \over a^3} + B + {2 \over \ell^2} \left[ 1 \pm \sqrt{1 + \ell^2
\left({A \over a^3} + B  - {\Lambda_{\rm b} \over 6} - {C \over a^4} \right)}
\right] \, ,
\end{equation}
where
\begin{equation}\label{ab}
A = {\rho_{0} a_0^3 \over 3 m^2} \, , \quad B = {\sigma \over 3 m^2} \, , \quad
\ell = {2 m^2 \over M^3} \, .
\end{equation}
Note that the four-dimensional Planck mass $m$ is related to the effective
Newton's constant on the brane as $m = 1/\sqrt{8\pi G}$. The two Planck masses
$M$ and $m$ define a new length scale $\ell = {2 m^2 / M^3} \sim cH_0^{-1}$ in
a braneworld which begins to accelerate at the current epoch \cite{DGP,ss02}.
(Other applications of braneworlds to the late-time acceleration of the
universe are discussed in
\cite{maeda03,lue-starkman,maartens04,brax04,vishwa03,alcaniz_pires,padilla04,brane_various}.)

The two signs in (\ref{hubble}) correspond to two branches of braneworld
models and are related to the two different ways in which
the brane can be embedded in the bulk. As shown in \cite{ss02}, the `$+$' sign in
(\ref{hubble}) corresponds to late time acceleration of the universe driven by
dark energy with an `effective' equation of state $w \geq -1$
(we call this model BRANE2) whereas
the `$-$' sign is associated with phantom-like behaviour $w \leq -1$ (this model
is called BRANE1).

Some limiting cases of (\ref{hubble}) will be of interest to the reader:
\begin{enumerate}

\item For $m = 0$ equation (\ref{hubble}) reduces to a Randall--Sundrum
universe \cite{BDL} described earlier in (\ref{eq:cosmolim}).

\item In the other extreme case when $M = 0$, extra-dimensional effects become
unimportant, and (\ref{hubble}) reduces to the LCDM model
\beq
H^2 (a) = {A \over a^3} + B \, .
\eeq

\item  Finally, when $\Lambda_{\rm b} = 0$ and $\sigma = 0$, (\ref{hubble}) reduces to the DGP
braneworld \cite{DGP}.

\end{enumerate}

Consider now the Braneworld with the `$-$' sign on the right-hand side of
(\ref{hubble}), \ie BRANE1. In this case the Hubble parameter on the brane can
be rewritten as:
$$
H^2 (a)  = {A \over a^3} + \Lambda_{\rm eff} \, .
$$
The term $\Lambda_{\rm eff}$ is composed of two terms, namely, a
constant $\Lambda$-term and a `screening term':
\beq\label{eq:screening}
\Lambda_{\rm eff} = \large\underbrace{ (B + {2 \over \ell^2}\large ) }
-  \underbrace{{2 \over \ell^2}\sqrt{1 + \ell^2
\left({A \over a^3} + B\right)  }}
 \, \nonumber
\eeq
\beq
\hskip 0.1cm \Downarrow \hskip 2.9cm \Downarrow
\eeq
\hskip 6.3cm {\Large{$\Lambda$}} \hskip 1.4cm Screening term

\bigskip

Since the screening term decreases with time, the value of {\bf the effective
cosmological constant $\Lambda_{\rm eff}$ increases}. Therefore the Braneworld
behaves just like a Phantom model ($w < -1$) but without Phantom's problems:
{\underline{no violation of the weak energy condition and no future
singularity}}. Indeed, from (\ref{eq:screening}) its clear that the universe
evolves to $\Lambda$CDM in the future (see also
\cite{lue-starkman,chiba00,kaloper01,caldwell02,cline04,star98,innes02,caldwell03,carroll03}
\cite{frampton03a,frampton03b,singh03,johri03,dabrowski03,odintsov04,nesseris04,kaloper04,srivastava04}
for discussions of related issues).

The fact that the Braneworld model (\ref{hubble}) can give rise to Phantom-like behaviour
can also be seen if we rewrite
(\ref{hubble}) in terms of the cosmological redshift `z' so that \cite{ss02}
\ber \label{eq:hubble_brane}
{H^2(z) \over H_0^2} &=& \Omega_{\rm m} (1\!+\!z)^3 + \Omega_\sigma +2 \Omega_l \pm
\nonumber\\
&&2 \sqrt{\Omega_l}\,
\sqrt{\Omega_{\rm m} (1\!+\!z )^3 + \Omega_\sigma + \Omega_\ell +
\Omega_{\Lambda_{\rm b}}} \, ,
\eer
where we have set the dark radiation term in (\ref{hubble}) to zero ($C=0$) and
\beq \label{eq:omegas}
\Omega_\ell = {1 \over \ell^2 H_0^2}~, ~\Omega_{\rm m} =  {\rho_{0m} \over 3 m^2 H_0^2}~, ~\Omega_\sigma
= {\sigma \over 3 m^2H_0^2}~,
~\Omega_{\Lambda_{\rm b}} = - {\Lambda_{\rm b} \over 6 H_0^2}~.
\eeq
The quantities $\Omega_i$ satisfy the constraint equation
\begin{equation} \label{omega-r2}
\Omega_{\rm m} + \Omega_\kappa + \Omega_\sigma \pm 2
\sqrt{\Omega_\ell}\, \sqrt{1 - \Omega_\kappa + \Omega_{\Lambda_{\rm b}}} = 1 \, ,
\end{equation}
where the $\pm$ sign in (\ref{omega-r2}) is commensurate with that in
(\ref{eq:hubble_brane}). For future reference we also note the cosmological
density associated with the four-dimensional cosmological constant
(\ref{eq:lam_RS}) in Randall--Sundram cosmology
\beq
\Omega_{\rm RS} = \frac{\Lambda_{\rm RS}}{3H_0^2} = \frac{\Omega_\sigma^2} {4
\Omega_\ell} - \Omega_{\Lambda_{\rm b}} \, .
\label{eq:Omega_RS}
\eeq
The expression for the
current value of the effective equation
of state is
\begin{equation}
w_0 = {2 q_0 - 1 \over 3 \left( 1 - \Omega_{\rm m} \right)} = - 1 \pm
{\Omega_{\rm m} \over 1 - \Omega_{\rm m}} \, {\sqrt{\Omega_\ell \over
\Omega_{\rm m} + \Omega_\sigma + \Omega_\ell + \Omega_{\Lambda_{\rm b}}}} \, ,
\label{eq:brane_state}
\end{equation}
and we immediately find that $w_0 \leq -1$ when we take
the lower sign in (\ref{eq:brane_state}), which corresponds to choosing one of
two possible embeddings of this braneworld in the higher dimensional bulk
(this is BRANE1).
(The second choice of embedding (BRANE2) gives $w_0 \geq -1$.)
It is important to note that all BRANE1 models 
have $w_0 \leq -1$ and $w(z) \simeq -0.5$ at $z \gg 1$.
They therefore successfully cross the `great divide' at
$w = -1$
(see page 26 of \cite{ss02} for a discussion of this issue).

\section{Transient Dark Energy}
\label{sec:DDE}

\begin{figure}[ht]
\centering
\includegraphics[width=9cm]{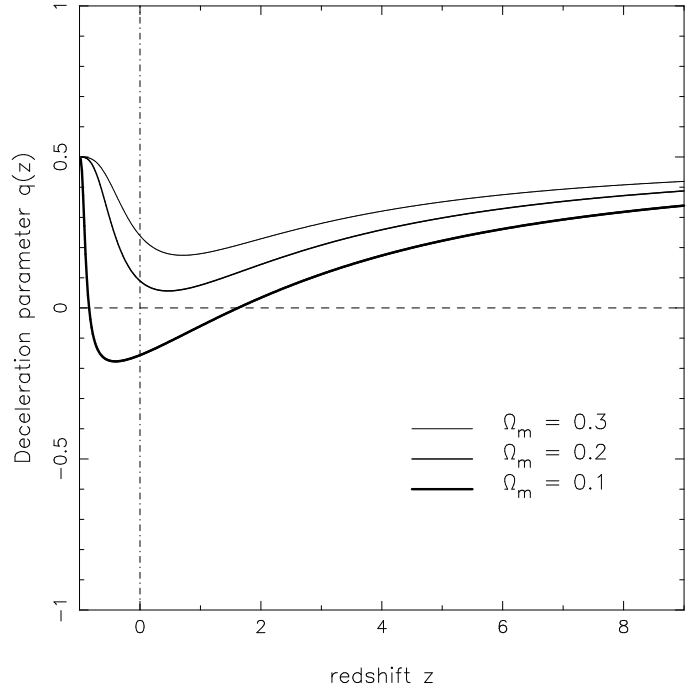}
\caption{\small The deceleration parameter $q=-{\ddot a}/aH^2$ is plotted as a
function of redshift for the BRANE2 model with the Randall--Sundrum constraint
$\Omega_\sigma = - 2 \sqrt{\Omega_\ell \Omega_{\Lambda_{\rm b}}}$. The vertical
(dot-dashed) line at $z = 0$ marks the present epoch, while the horizontal
(dashed) line at $q = 0$ corresponds to a Milne universe [$a(t) \propto t$]
which neither accelerates nor decelerates. The case $q = 0.5$ corresponds to
matter dominated expansion. Note that the universe ceases to accelerate and
becomes matter dominated in the past {\em as well as in the future}. Figure
courtesy of  \cite{ss02}.} \label{fig:decel_plot}
\end{figure}

Braneworld cosmology has the interesting property that it
allows the acceleration of the universe to be a transient phenomenon.
(In \cite{ss02} these models were referred to as
`Disappearing Dark Energy' (DDE).)
Indeed, by setting the effective four dimensional cosmological constant
(\ref{eq:lam_RS}) to zero so that $\Omega_{\rm RS} = 0$ in (\ref{eq:Omega_RS}) we get
\beq
\Omega_\sigma = \pm 2 \sqrt{\Omega_\ell \Omega_{\Lambda_{\rm b}}}~.
\label{eq:RSconstraint}
\eeq
Selecting the lower sign in (\ref{eq:RSconstraint}) and substituting
$\Omega_\sigma = - 2 \sqrt{\Omega_\ell \Omega_{\Lambda_{\rm b}}}$ in
(\ref{eq:hubble_brane}) we find that\footnote{Note that the case
$\Omega_\sigma = 2 \sqrt{\Omega_\ell \Omega_{\Lambda_{\rm b}}}$
leads to transient acceleration with $\Omega_m > 1$ which is unrealistic \cite{ss02}.}
\begin{equation} \label{negative}
\left\lbrack{H(z = -1) \over H_0}\right\rbrack^2 = 2 \sqrt{\Omega_\ell} \left(
\sqrt{\Omega_\ell} - \sqrt{\Omega_{\Lambda_{\rm b}}} \pm
\left|\sqrt{\Omega_\ell} - \sqrt{\Omega_{\Lambda_{\rm b}}} \right|\, \right) \,
.
\end{equation}
which {\em vanishes\/} for the upper sign when $\Omega_\ell \le
\Omega_{\Lambda_{\rm b}}$. This implies that, since the effective four
dimensional cosmological constant vanishes,
the universe reverts to matter dominated
expansion in the future: \\
$H^2(z\to -1) \simeq \Omega_m(1+z)^3\to 0$. The current acceleration of the
universe is therefore a transient phase in this model, as illustrated in
figure~\ref{fig:decel_plot}.

It is well known that an eternally
accelerating universe is endowed with a cosmological event horizon which
prevents the construction of a conventional S-matrix describing particle
interactions within the framework of string or M-theory \cite{horizon}.
A transiently accelerating braneworld may therefore help reconcile
string/M-theory with accelerating cosmology, which is an attractive feature of
this braneworld cosmology.

\section{Quiescent Future Singularity}

As remarked earlier, an interesting property of the Braneworld (\ref{hubble})
is that the late-time expansion of the
universe can culminate in a `Quiescent' singularity at which the density,
pressure and Hubble parameter remain finite while the deceleration parameter
and geometrical invariants constructed from the Riemann tensor diverge \cite{ss02b}.

In order to appreciate the origin of such unusual future singularities in the
braneworld, consider  again the expansion law (\ref{eq:hubble_brane}). For
simplicity, we shall only discuss the solution corresponding to the `+' sign in
(\ref{eq:hubble_brane}) (called BRANE2 in \cite{ss02}).

A necessary and sufficient condition for the existence of a Quiescent
singularity is for the inequality
\begin{equation} \label{constraint}
\Omega_\sigma + \Omega_\ell + \Omega_{\Lambda_{\rm b}}  < 0 \, ,
\end{equation}
to be satisfied \cite{ss02b}. In this case
the expression under the
square root of (\ref{eq:hubble_brane}) becomes zero at a suitably late
time, and the cosmological solution {\em cannot be extended beyond this
time}.
It can be shown that the scale factor $a(t)$ and its first time derivative remain
finite, while all the higher time derivatives of $a(t)$ {\em tend to infinity} as the
singularity is approached.

\begin{figure}[ht]
\centering
\includegraphics[width=9cm]{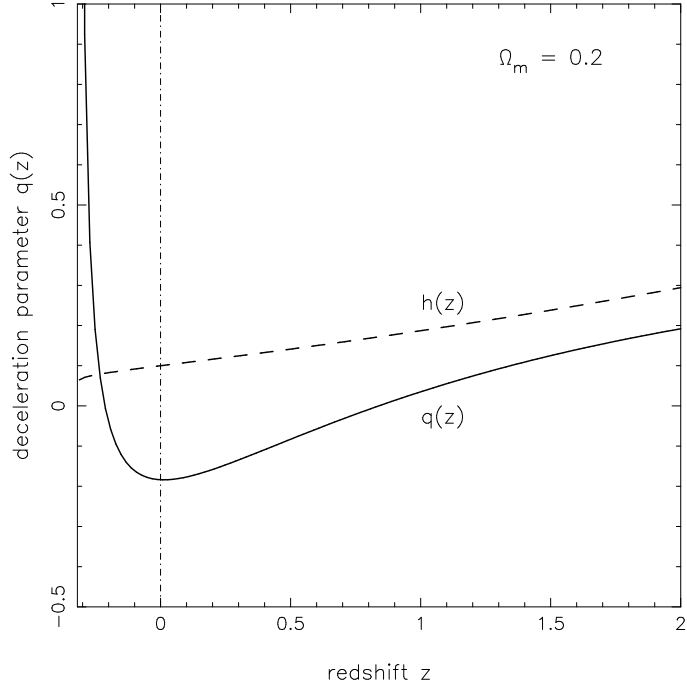}
\caption{\small The deceleration parameter (solid line) is shown for a
braneworld model with $\Omega_{\rm m} = 0.2$, $\Omega_\ell = 0.4$,
$\Omega_\Lambda = \Omega_\kappa = 0$, and $\Omega_\sigma$ determined from
(\ref{omega-r2}). We find that $q(z) \to 0.5$ for $z \gg 1$ while $q(z) \to
\infty$ as $z \to -0.312779$... Currently $q_0 < 0$, which indicates that the
universe is accelerating. Also shown is the dimensionless Hubble parameter
$h(z) = 0.1\times H(z)/H_0$ (dashed line) which remains finite in this model.
The vertical line at
$z = 0$ shows the present epoch. Figure courtesy of \cite{ss02b}.}
\label{fig:decel}
\end{figure}

The limiting redshift, $z_s = a_0/a(z_s) - 1$, at which the braneworld
becomes singular is given by
\begin{equation}
z_s = \left (-\frac{\Omega_\sigma + \Omega_\ell + \Omega_{\Lambda_{\rm b}}} {\Omega_{\rm
m}}\right )^{1/3} - 1 \, .
\end{equation}
The time of occurance of the singularity (measured from the present moment)
can easily be determined from
\beq
T_s = t(z=z_s) - t(z=0) = \int_{z_s}^0\frac{dz}{(1+z)H(z)},
\eeq
where $H(z)$ is given by (\ref{eq:hubble_brane}).
In Fig.~\ref{fig:decel} we show a specific braneworld model having $\Omega_{\rm
m} = 0.2$, $\Omega_\ell = 0.4$, $\Omega_{\Lambda_{\rm b}} = \Omega_\kappa = 0$. In
keeping with observations of high redshift supernovae our model universe is
currently accelerating \cite{sn}, but will become singular at $z_s \simeq -0.3
\Rightarrow a(z_s) \simeq 1.4\times a_0$,
i.e. after $T_s \simeq 4.5~ h_{100}^{-1}$ Gyr ($h_{100} = H_0/100$ km/sec/Mpc).
Figure~\ref{fig:decel} demonstrates that
the deceleration parameter becomes singular as $z_s$ is approached:
\begin{equation}
q = -\frac{\ddot a}{aH^2} \equiv \frac{H'}{H} (1+z) - 1 \, ; \quad \lim_{z \to
z_s} q(z) \to \infty \, ,
\end{equation}
while the Hubble parameter remains finite:
\begin{equation}
{H^2(z_s) \over H_0^2} = \Omega_\ell - \Omega_{\Lambda_{\rm b}} \, .
\end{equation}

The possible presence of such `Quiescent' singularity in Braneworld theory can
also be seen from the following consideration. As shown in \cite{ss02}, the
action (\ref{action}) leads to the bulk being described by the usual Einstein
equation with cosmological constant:
\begin{equation} \label{bulk}
{\cal G}_{ab} + \Lambda_{\rm b} g_{ab} = 0 \, ,
\end{equation}
while the field equation on the brane is
\begin{equation} \label{brane}
m^2 G_{ab} + \sigma h_{ab} = \tau_{ab} + \underline{M^3 \left(K_{ab} -
h_{ab} K \right)} \, .
\end{equation}
Here, $\tau_{ab}$ is the stress-energy tensor on the brane. A divergent form
for the extrinsic curvature (underlined term) caused by a singular embedding of
the brane in the bulk will, through (\ref{brane}), lead to a singular value for
the Einstein tensor $G_{ab}$ even though all components of the stress-energy
tensor on the brane remain finite\,!

More generally it is well known that Quiescent singularites occur when the original
equations of motion are non-linear with respect to the highest derivative. They
have earlier been discussed in the context of Einstein gravity with the
conformal anomaly \cite{wfhh}. (This result is not surprising in view of the
formal similarity between braneworld theory and GR-based models with the
conformal anomaly, discussed in \cite{ss02}.) `Determinant singularities'
having a similar structure and properties are known to arise in the anisotropic
Bianchi~I model containing a dilaton coupled to a Gauss--Bonnet term in the
action \cite{topor} and also in other cosmological models \cite{kamen04,barrow04}.
For instance \cite{kamen04} refer to singularities in which the deceleration
parameter tends to infinity as the `Big Brake' while in
\cite{barrow04} they are called `sudden' singularities (see also \cite{cotsakis04}).

We should also emphasise that both the past and
future quiescent singularities occur for a wide range
in parameter space and might provide an interesting alternative to the `big
bang'/`big crunch' singularities of general relativity.

\section{Loitering Braneworld Models}

\begin{figure}[ht]
\centering
\resizebox{5.5in}{!}{\rotatebox{-90}{
\includegraphics{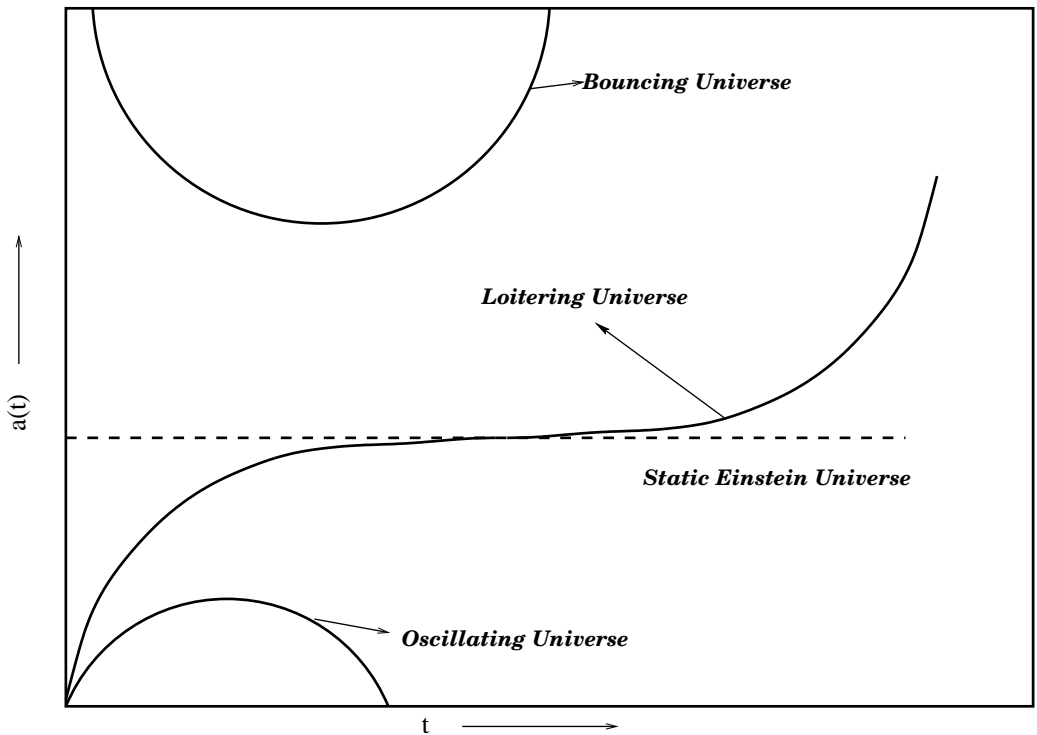}
}}
\caption{\small Dynamical possibilities in a closed FRW universe are
essentially the same as in the { \em spatially flat brane} (\ref{eq:loit1}) and
are summarised above. Figure courtesy of  \cite{ss00}. } \label{fig:loiter}
\end{figure}

An interesting aspect of the Braneworld models (\ref{hubble}) is that they
allow for the existence of an early {\em loitering\/} epoch \cite{ss04}.
Loitering is characterized by the fact that the Hubble parameter dips in value
over a narrow redshift range referred to as the `loitering epoch'. During
loitering density perturbations are expected to grow rapidly. In addition,
since the expansion of the universe slows down, its age near loitering
dramatically increases. An early epoch of loitering is expected to boost the
formation of high redshift gravitationally bound systems such as $10^9 M_\odot$
black holes at $z \sim 6$ and lower-mass black holes and/or Population III
stars at $z > 10$, whose existence could be problematic within the LCDM
scenario.

To demonstrate the possibility of loitering note that for
large values of the `dark radiation' term
$C/a^4$, when $C< 0$ and $\ell^2 {|C|/a^4} \gg 1$, equation
(\ref{hubble}) acquires the form
\begin{equation}
H^2(a) \approx {A \over a^3} + B \pm {2 \sqrt{- C} \over \ell a^2} \, .
\label{eq:loit1}
\end{equation}
Equation (\ref{eq:loit1}) bears a formal resemblance to the Hubble parameter
in standard GR
\beq
H^2 = \frac{8\pi G}{3}\frac{\rho_{0} a_0^3}{a^3} + \frac{\Lambda}{3} -
\frac{\kappa}{a^2} \, . \label{eq:frw}
\eeq
Remarkably,
the role of the spatial curvature in (\ref{eq:frw}) is played by
the dark-radiation term in (\ref{eq:loit1});
consequently, a spatially open universe is mimicked by
the BRANE2 model whereas a closed universe is mimicked by BRANE1.

\begin{figure}[ht]
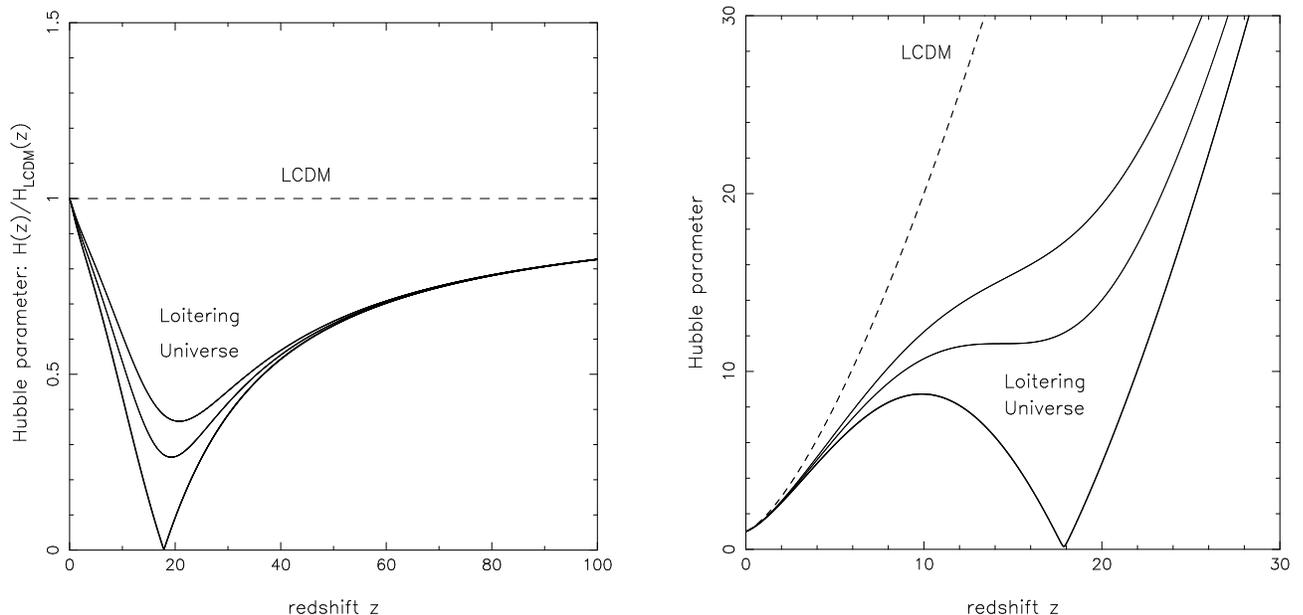

\centering
$\begin{array}{c@{\hspace{0.4in}}c}
\multicolumn{1}{l}{\mbox{}} &
\multicolumn{1}{l}{\mbox{}} \\ [0.1cm]
\includegraphics[width=8cm]{hub_loiter2.epsi} &
\includegraphics[width=8cm]{hub_loiter1.epsi}
\end{array}$
\caption{\small The Hubble parameter for three different universes all of which
loiter at $z_{\rm
loit} \simeq 18$. Parameter values are $\Omega_{\rm m} = 0.3$, $\Omega_C =
8.0$, $\Omega_\ell = 3.0$, and $\Omega_{\Lambda_{\rm b}}/10^5 = 6,\, 4.5,\,
3.4$ (solid lines, from top to bottom). The left panel shows the Hubble
parameter with respect to the LCDM value while, in the right panel, the LCDM
(dashed) and loitering (solid)
Hubble parameters are shown separately. Figure courtesy of  \cite{ss04}.} \label{fig:hubble}
\end{figure}

Loitering solutions in standard cosmology were first found for a closed
FRW model with a cosmological constant in \cite{lemaitre}.
Loitering in more general situations was discussed in \cite{sfs92,robert,polarski,ss04}.
An intriguing aspect of braneworld models is that, since a spatially flat braneworld
can behave dynamically just like
a closed universe, loitering may be realised within a spatially flat setting and therefore
be consistent both with Inflationary cosmology and recent CMB results \cite{wmap}.
Indeed, loitering solutions to (\ref{hubble}) \& (\ref{eq:loit1}) can easily be found
by requiring that the
loitering condition $dH/dz = 0$ is satisfied.
This allows us to determine the `loitering redshift' \cite{ss04}
\beq
1\!+\!z_{\rm loit} \simeq \frac{4}{3} \,
\frac{\sqrt{\Omega_C\Omega_\ell}}{\Omega_{\rm m}} \, . \label{eq:loit}
\eeq
From this expression we find that the universe will loiter at a large redshift
$z_{\rm loit} \gg 1$ provided $\Omega_C \Omega_\ell \gg \Omega_{\rm m}^2$.
Since $\Omega_{\rm m}^2 \ll 1$, loitering at large redshifts is
not difficult to achieve.

One should note here that a closed FRW universe with a cosmological constant
can loiter only at rather small redshifts: $z_{\rm loit} \leq 2$ for
$\Omega_{\rm m} \geq 0.1$ \cite{sfs92}. Therefore Braneworld cosmology
is endowed with two fundamentally new attributes:
(i) it permits loitering in a spatially flat universe and (ii)
it allows the loitering redshift to be large.
Neither of these possibilities is allowed in standard Einsteinian cosmology.

The Hubble parameter at loitering is given by the approximate
expression \cite{ss04}
\beq
{H^2(z_{\rm loit}) \over H_0^2} \simeq \Omega_\sigma -
\frac{32}{27}\frac{(\Omega_C\Omega_\ell)^{3/2}}{\Omega_{\rm m}^2} \, .
\label{eq:hub_loiter}
\eeq

Examples of a loitering model are shown in Fig.~\ref{fig:hubble}, where the
Hubble parameter of a universe which loitered at $z \simeq 18$ is plotted
against the redshift.
The right hand panel of figure \ref{fig:hubble} illustrates the fact
that the loitering universe
can show a variety of interesting behaviour:
(i) top curve, $H(z)$ is monotonically increasing and
$H'(z) \simeq {\rm constant}$ in the loitering interval;
(ii) middle curve, $H(z)$ appears to have an inflexion point
($H' \simeq 0, H'' \simeq 0$) during loitering;
(iii) lower curve, $H(z)$ has both a maximum and a mininimum, the latter occuring
in the loitering regime.

\begin{figure}[ht]
\centering
\includegraphics[width=9cm]{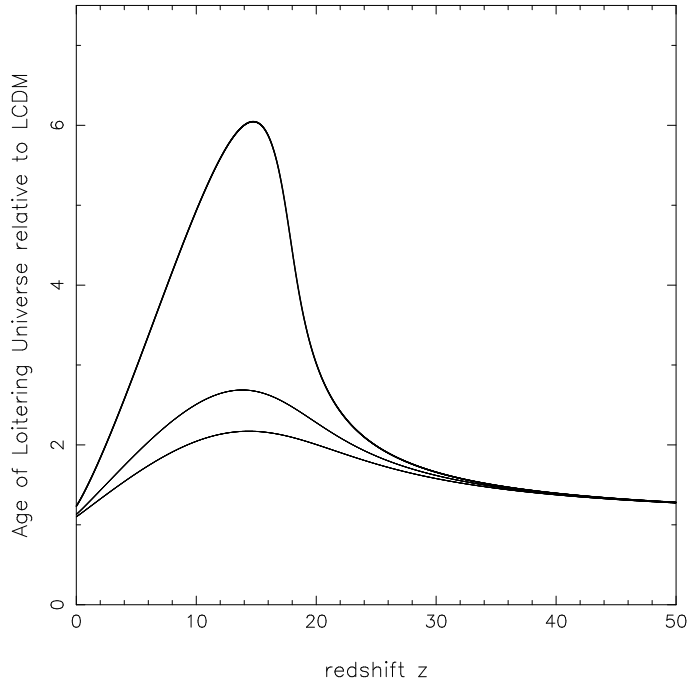}
\caption{\small
The age of three loitering models is shown relative to the age
in LCDM (model parameters are the same as in Fig.~\ref{fig:hubble}). Note that
the age of the universe near loitering ($z_{\rm loit} \sim 18$) is
significantly greater than that in LCDM although, at the present epoch, the
difference in ages between the two models is relatively small.
Figure courtesy of  \cite{ss04}.}

\label{fig:age}
\end{figure}

A loitering universe could have several important cosmological consequences,
one of them being that
the age of the universe during loitering {\em increases\/}, as shown in
Fig.~\ref{fig:age}. The reason for the age increase rests in the expression
\beq
t(z) = \int_z^\infty \frac{dz'}{(1+z')H(z')} \, . \label{eq:age}
\eeq
which shows that a lower value of $H(z)$ close to loitering will boost the age of the
universe at that epoch.
An important consequence of having a larger age of the universe at $z \sim 20$
(or so) is that astrophysical processes at these redshifts have more time in
which to develop. This is especially important for gravitational instability
which forms gravitationally bound systems from the extremely tiny fluctuations
existing at the epoch of last scattering. Thus, an early loitering epoch may be
conducive to the formation of Population~III stars and low-mass black holes at
$z \sim 17$ and also of $\sim 10^9 M_\odot$ black holes at lower redshifts ($z
\sim 6$) whose existence could be problematic within the LCDM scenario
\cite{richards03}.
(Note that
age of a LCDM universe at $z \gg 1$ is $t(z) \simeq (2/3H_0\sqrt{\Omega_{\rm
m}}) (1+z)^{-3/2} = 5.38 \times 10^8 (1+z/10)^{-3/2} $ years for $\Omega_{\rm
m} = 0.3, h = 0.7$.)

Another important property of a loitering universe is that it
can alter the reionization properties of the inter-galactic medium at 
moderate redshifts $z \lleq z_{\rm loiter}$. 
One should note in this context that a major
surprise emerging from the WMAP experiment was the discovery of a large
optical depth to reionization $\tau = 0.17 \pm 0.06$ \cite{wmap}, which,
when translated within
the framework of LCDM (assuming
instantaneous reionization), implies a rather early epoch for
reionization $z_{\rm reion} \simeq 17 \pm 5$.

In order to appreciate how loitering might alter these conclusions 
consider the following expression for the
optical depth to a redshift $z_{\rm reion}$ 
\beq
\tau(z_{reion}) = c\int_0^{z_{reion}}\frac{n_e(z)\sigma_T ~dz}{(1+z)H(z)}
\eeq
where $n_e$ is the electron density and $\sigma_T$ is the Thompson cross-section
describing scattering between electrons and CMB photons.
During loitering $H(z)$ drops below its value in LCDM, therefore
$z_{\rm reion}\vert_{\rm loitering} < z_{\rm reion}\vert_{\rm LCDM}$.
For instance the redshift of reionization drops to
$z_{reion} \leq 12$ for the loitering models shown in figure \ref{fig:hubble}.
Loitering decreases
the redshift of reionization and increases the
age of the universe thereby
alleviating the
existing tension between the high redshift universe and dark energy cosmology
\cite{ss04}.

\subsection{The effective equation of state in a loitering Braneworld.}

The deceleration parameter $q$ and the effective equation of state $w$ in our
loitering model are given by the expressions
\ber
q(z) &=& \frac{H'(z)}{H(z)} (1+z) - 1 \, ,\nonumber\\
w(z) &=& {2 q(z) - 1 \over 3 \left[ 1 - \Omega_{\rm m}(z) \right] } \, ,
\label{eq:state0}
\eer
where $H(z)$ is determined from (\ref{eq:hubble_brane}).
The current values of these quantities are
\ber
q_0 &=& \frac{3}{2}\Omega_{\rm m}\left [1 -
\frac{\sqrt{\Omega_\ell}}{\sqrt{\sum_i\Omega_i}}
\left(1 + \frac{4}{3}\frac{\Omega_C}{\Omega_{\rm m}}\right)\right ] - 1 \, ,\nonumber\\
w_0 &=& -1 - \frac{\Omega_{\rm m}}{(1-\Omega_{\rm m})}\frac{\sqrt{\Omega_\ell}}
{\sqrt{\sum_i\Omega_i}} \left(1 + \frac{4}{3}\frac{\Omega_C}{\Omega_{\rm m}}
\right) \, , \label{eq:w_0}
\eer
where $\sum_i\Omega_i = \Omega_{\rm m} + \Omega_C + \Omega_\ell + \Omega_\sigma
+ \Omega_{\Lambda_{\rm b}} = \left( \sqrt{\Omega_\ell} + \sqrt{1 +
\Omega_{\Lambda_{\rm b}} + \Omega_C} \right)^2$. From Eq.~(\ref{eq:w_0}) we
find that $w_0 < -1$ if $\Omega_C \geq 0$; in other words, our loitering
universe has a phantom-like effective equation of state (EOS). (In particular,
for the loitering models shown in Fig.~\ref{fig:hubble}, we have $w_0 =
-1.035\,$, $-1.04\,$, $-1.047$ (top to bottom), all of which are in excellent
agreement with recent observations \cite{alam04,seljak04}.)

An interesting consequence of the loitering braneworld is that the matter
density $\Omega_{\rm m} = 8\pi G \rho_{\rm m} /3H^2$ {\em exceeds\/} unity at
some time in the past. This follows immediately from the fact that, since the
value of $H(z)$ in the loitering braneworld model is {\em smaller\/} than its
counterpart in LCDM, the value of $\Omega_{\rm m}(z)$ is larger than its
counterpart in LCDM. One important consequence of this behaviour is that, as
expected from (\ref{eq:w_0}), the effective equation of state (EOS) blows up
precisely when $\Omega_{\rm m}=1$.

\begin{figure}[ht]
\centering
\includegraphics[width=9cm]{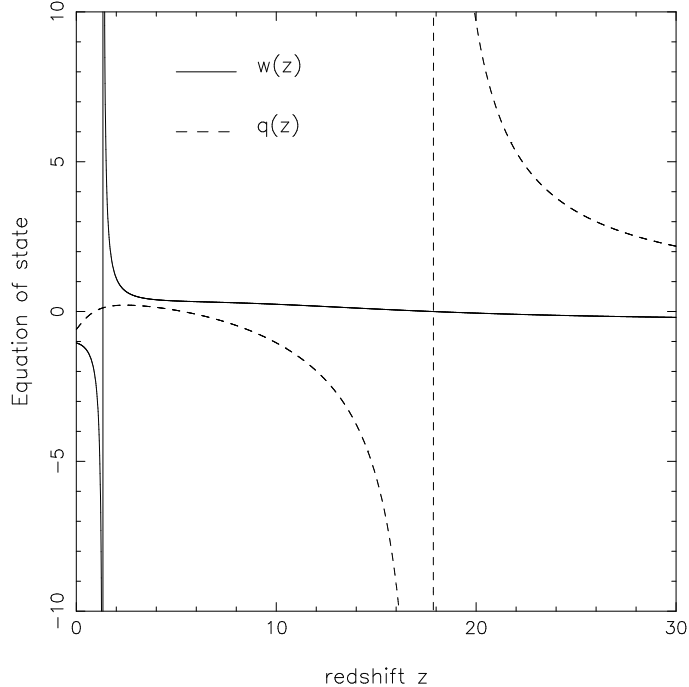}
\caption{\small The effective equation of state of dark energy (solid) and the
deceleration parameter (dashed) are shown for a universe which loitered at $z
\simeq 18$. Note that the effective equation of state of dark energy becomes
infinite at low redshifts when $\Omega_{\rm m}(z) = 1$. However, this behaviour
is not reflected in the deceleration parameter, which becomes large only near
the loitering redshift when $H \simeq 0$. Figure courtesy of  \cite{ss04}. } \label{fig:state}
\end{figure}

In Fig.~\ref{fig:state}, we show that, in
contrast to the singular behaviour of the EOS, the deceleration parameter
remains finite and well behaved even as $w \to \infty$. Note that the {\em finite
behaviour} of $q(z)$ when $\Omega_{\rm m}(z) = 1$
reflects the fact that the EOS for the braneworld is an
{\em effective\/} quantity and not a real physical property of the theory.
This is an important point with broader ramifications since it concerns
dark energy models constructed from modifications to the `gravity sector'
of the theory of which Braneworld models are but one example
(others being: scalar-tensor gravity, Cardassian approximation,
$R+f(R)$ gravity, k-essence, to name but a few).
Let me therefore dwell on this point in some more detail.

Within the framework of general relativity the
rate of expansion  of a FRW universe and its acceleration are
described by the pair of equations
\ber\label{eq:acc}
H^2 &=& \frac{8\pi G}{3}\sum_i \rho_i - \frac{k}{a^2}~,\nonumber\\
\frac{\ddot a}{a} &=& -\frac{4\pi G}{3}\sum_i (\rho_i
+ 3p_i),
\eer
where the summation is over all matter fields contributing to the
dynamics of the universe.
Within this framework,
and in the absence of spatial curvature, the energy density and
pressure of dark energy can be defined as:
\ber\label{eq:energy}
\rho_{\rm DE} &=& \rho_{\rm critical} - \rho_{\rm m} =
\frac{3H^2}{8\pi G}\left( 1 - \Omega_{\rm m} \right)~,\nonumber\\
p_{\rm DE} &=& \frac{H^2}{4\pi G} \left(q - \frac{1}{2} \right)~,
\eer
where $\rho_{\rm critical} = 3H^2/8\pi G$ is the critical density associated
with a FRW universe and $q=-{\ddot a}/{aH^2}$ is the deceleration parameter. An
important consequence of using (\ref{eq:acc}) \& (\ref{eq:energy}) is that the
ratio $w_{DE}\equiv p_{DE}/\rho_{DE}$ can be expressed in terms of the
deceleration parameter
\beq\label{eq:state}
w_{\rm eff}(x) = {2 q(x) - 1 \over 3 \left( 1 - \Omega_{\rm m}(x) \right)}
\equiv \frac{(2 x /3) \ d \ {\rm ln}H \ / \ dx - 1}{1 \ - \ (H_0/H)^2
\Omega_m \ x^3}\,\,~,~~~ x = 1 + z ~.
\eeq
\ie we recover equation (\ref{eq:state0}) which had earlier been used to determine
the equation of state in a loitering universe. Since this
derivation of $w_{\rm eff}(z)$ uses the general relativistic formulation
it can be expected to recover standard results for GR-based dark energy
models such as
LCDM
and quintessence.
However the same cannot be said of braneworld
models since the Hubble parameter for the latter contains interaction
terms between matter and dark energy (see for instance equation
(\ref{eq:hubble_brane})) and therefore does not
subscribe to the general relativistic format (\ref{eq:acc}).
``One can however extend the above definition of $w_{\rm eff}$ to
non-Einsteinian theories by {\em defining} dark energy density to be
the remainder term after one subtracts the matter density from the
critical density in the Einstein equations.  It should be emphasised
that, according to this prescription all interaction terms between
matter and dark energy (such terms arise in scalar-tensor and brane
models) are attributed {\em solely} to dark energy. Therefore $w_{\rm
eff}(z)$ defined according to (\ref{eq:state}) is an {\em effective
equation of state} in these models and not a fundamental physical
entity (as it is in LCDM, for instance)'' \cite{alam}.
The fact that the EOS is indeed an effective equation of state
and not a fundamental physical
quantity is illustrated by figure \ref{fig:state} which shows that $w_{\rm eff}(z)$
{\em diverges} when $\Omega_m(z) \simeq 1$ and that this divergence is
not reflected in other characteristics of expansion such as the
deceleration parameter.
\footnote{It should also be stressed that the propagation velocity
of small inhomogeneities in dark energy is generically neither
$\sqrt{w_{DE}}$, nor $\sqrt{dp_{DE} / d\rho_{DE}}$. Therefore although
$w(z)$ is an important physical quantity it does not provide us with
an exhaustive description of dark energy and its use as a diagnostic
should be treated with some caution, as emphasised in \cite{alam}.}

\subsection{The Statefinder diagnostic and Dark Energy}

As demonstrated above, the equation of state is not a fundamental
physical quantity for dark energy models based
either on non-Einsteinian gravity (scalar-tensor theories, Cardassian approximation,
k-essence, R + f(R) gravity, etc.) or on Einstein gravity in more than 3+1 dimensions,
or indeed in cosmological models with explicit interaction terms between dark matter
and dark energy.
In all these cases it is useful to supplement the effective equation of state
by other geometrical variables, which are `model independent', in the sense that
they do not depend upon an underlying theory of gravity in an explicit way,
but can be constructed solely from the expansion factor and its derivatives,
and therefore would be expected to apply to all metric theories of gravity
defined on a space-time manifold. In all such theories the universe can be
expected to be characterised by the following general form for the expansion
factor:
\beq\label{eq:taylor0}
a(t) = a(t_0) + {\dot a}\big\vert_0 (t-t_0) +
\frac{\ddot{a}\big\vert_0}{2} (t-t_0)^2 +
\frac{\atridot\big\vert_0}{6} (t-t_0)^3 + ...~.
\eeq

Usually dark energy models such as quiessence, quintessence,
k-essence, braneworld models, Chaplygin gas etc. give rise to families
of curves $a(t)$ having quite different properties.  Since we know
that the acceleration of the universe is a fairly recent phenomenon
we can, in principle, confine our
attention to small values of $\vert t-t_0\vert$ in (\ref{eq:taylor0}).

Accordingly, as dicussed in \cite{sssa02,alam}, a new diagnostic of dark energy
called `statefinder' can be constructed using both the second and third
derivatives of the expansion factor (see also \cite{cn98,visser04,linder04}).
The statefinder pair $\statei$, defines two new cosmological
parameters (in addition to $H$ and $q$):
\ber\label{eq:statefinder1}
r &\equiv& \frac{\atridot}{a H^3} = 1 + \frac{9w}{2} \omx(1+w) -
\frac{3}{2} \omx \frac{\dot{w}}{H} \,\,,\\
s &\equiv& \frac{r-1}{3(q-1/2)} = 1+w-\frac{1}{3} \frac{\dot{w}}{wH}\,\,.
\eer

\begin{figure}[ht]
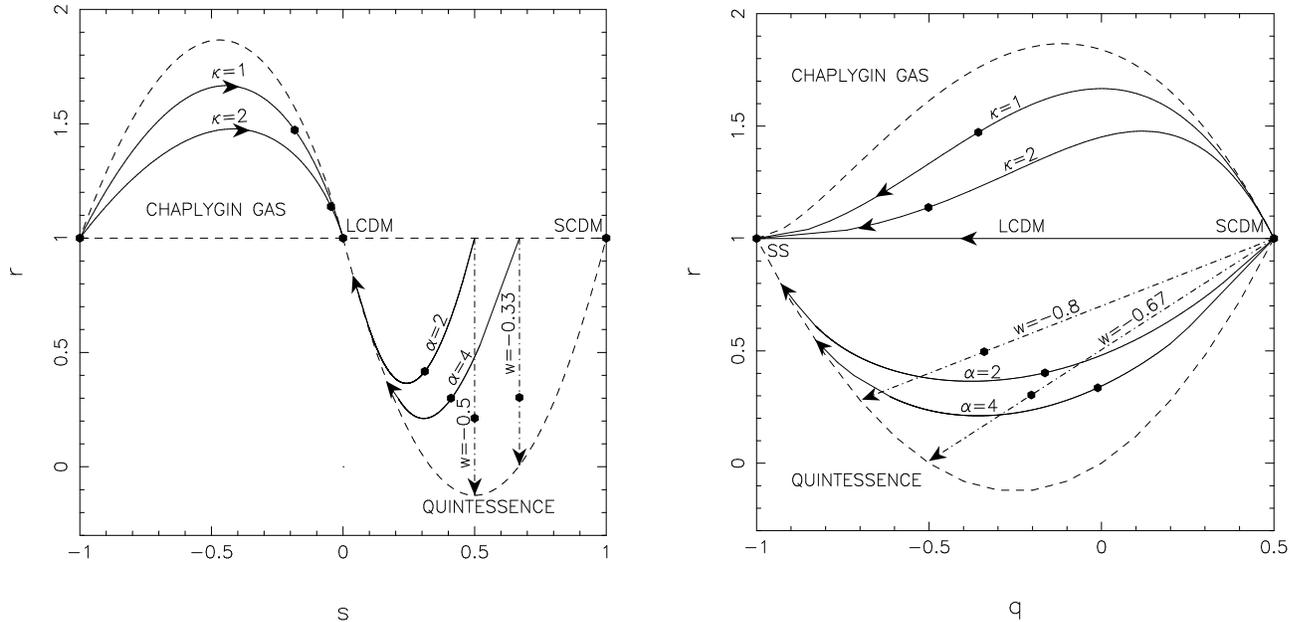

\centering
$\begin{array}{c@{\hspace{0.4in}}c}
\multicolumn{1}{l}{\mbox{}} &
\multicolumn{1}{l}{\mbox{}} \\ [0.1cm]
\includegraphics[width=8cm]{evolve_rs.epsi} &
\includegraphics[width=8cm]{evolve_qr.epsi}
\end{array}$
\caption{\small
The left panel (a) shows the time evolution of the statefinder pair
$\lbrace r,s \rbrace$ for quintessence models and the Chaplygin gas.
Quintessence models lie to the right of the LCDM fixed point
($r=1,s=0$) (solid lines represent tracker potentials
$V=V_0/\phi^{\alpha}$, dot-dashed lines representing quiessence with
constant equation of state $w$).  For quiessence models, $s$ remains
constant at $1+w$ while $r$ declines asymptotically to
$1+\frac{9}{2}w(1+w)$. For tracker models, $s$ monotonically decreases
to zero whereas $r$ first decreases from unity to a minimum value,
then rises to unity. These models tend to approach the LCDM fixed
point ($r=1, s=0$) from the right at $t \to \infty$.  Chaplygin gas
models (solid lines) lie to the left of the LCDM fixed point.
The right panel (b) shows the
time evolution of the pair $\lbrace r,q\rbrace$, where $q$ is the
deceleration parameter.
The solid
line, which corresponds to the time evolution of the LCDM model,
divides the $r-q$ plane into two halves. The upper half is occupied by
Chaplygin gas models, while the lower half contains quintessence
models. All models diverge at the same point in the past ($r=1,q=0.5$)
which corresponds to a matter dominated universe (SCDM), and converge
to the same point in the future ($r=1,q=-1$) which corresponds to the
steady state model (SS) -- the de Sitter expansion (LCDM $\to$ SS as
$t \to \infty$ and $\om \to 0$).  The dark dots on the curves show
current values $\lbrace r_0, s_0\rbrace$ (left) and $\lbrace r_0,
q_0\rbrace$ (right) for different dark energy models. In all models,
$\om = 0.3$ at the current epoch.  In both panels quiessence is shown
as dot-dashed while dashed lines mark envelopes for Chaplygin gas
(upper) and quintessence (lower).  Figure courtesy of  \cite{alam}.}
\label{fig:evolve}
\end{figure}

In figure \ref{fig:evolve}
we show the statefinder pair for several dark energy models including
LCDM, quintessence and the Chaplygin gas.
For the sake of completeness we say a few words about each of these models
below. (Note that the ensuing discussion is largely based on \cite{sssa02,alam}
and the reader
should refer to these papers for more details.)

\begin{figure}[ht]
\centering
\includegraphics[width=9cm]{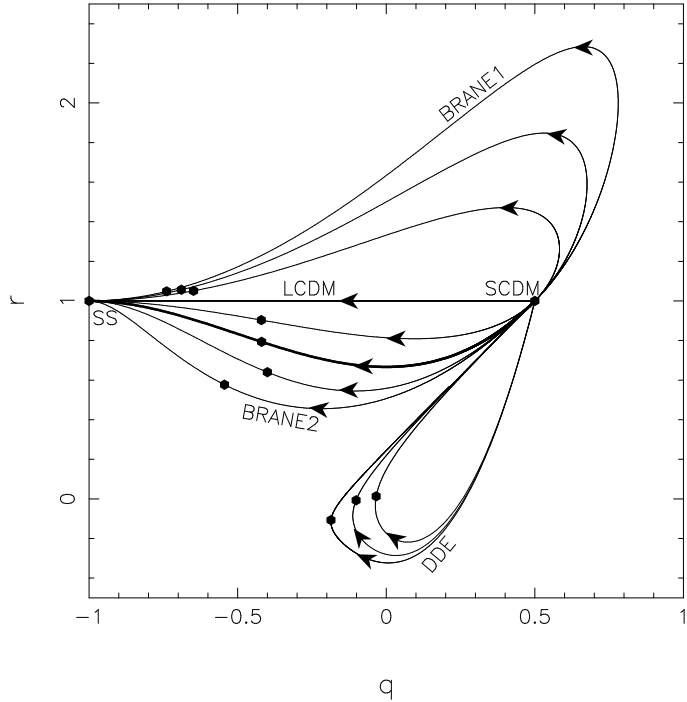}
\caption{\small
Trajectories in the statefinder plane $\statej$ for the {\bf Braneworld
models} discussed in (\ref{eq:hubble_brane}).  BRANE1 models have $w
\leq -1$ generically, whereas BRANE2 models have $w \geq -1$.  The
closed loop represents `Disappearing Dark Energy' (DDE) (discussed in
section \ref{sec:DDE})
for which the acceleration of the universe
is a transient phenomenon.
The thick solid curve in BRANE2 corresponds to the
DGP model \cite{DGP}.
Dark dots indicate the current
value of $\statej$ for the models.  All models are in reasonable
agreement with current supernova data.
One can see that the statefinder plane $\statej$ easily helps distinguish
the three Braneworld models from each other.
Figure courtesy of  \cite{alam}.
}
\label{fig:brane}
\end{figure}

\begin{itemize}

\item In {\bf Quiessence}
the equation
of state of dark energy is a constant so that
\beq\label{eq:quiessence}
H(z) = H_0 \left\lbrack \Omega_{\rm m}(1+z)^3 +
\omx(1+z)^{3(1+w)}\right\rbrack^{1/2}~.
\eeq
where $w$ = constant.  Important
examples of quiessence include: LCDM ($w = -1$), a network of non-interacting cosmic
strings ($w = -1/3$), domain walls ($w = -2/3$) and phantom models
($w < -1$). Quiessence in a
FRW universe can also be produced by a scalar field (quintessence)
which has the potential $V(\phi)\propto \sinh ^{{-2(1+w)\over
    w}}(C\phi+D)$, with appropriately chosen values of $C$ and $D$
    \cite{ss00,um00,sssa02}.

\item
The density and pressure of {\bf Quintessence}
are given by
\beq\label{eq:scalar}
\rho_{\phi} = \half \dot{\phi}^2+V(\phi), ~~
p_{\phi} = \half \dot{\phi}^2-V(\phi)~,
\eeq
while evolution of the Quintessence field $\phi$ is governed by the equation of motion
\beq
\ddot{\phi}+3 H \dot{\phi} +\frac{dV}{d\phi} = 0~,
\eeq
where
\beq\label{eq:kinessence}
H^2 = \frac{8\pi G}{3}\left\lbrack \rho_{0m}(1+z)^3 + \half{\dot\phi}^2+
V(\phi)\right\rbrack.
\eeq
It is clear from (\ref{eq:scalar}) that $w < -1/3$ provided
$\dot{\phi}^2 < V(\phi)$. Models with this property can lead to an
accelerating universe at late times. An important subclass of
Quintessence models displays the so-called `tracker' behaviour during
which the ratio of the scalar field energy density to that of the
matter/radiation background changes very slowly over a substantial
period of time.  Models belonging to this class satisfy $V^{\prime
  \prime} V/(V^{\prime})^2 \geq 1$ and approach a common evolutionary
  `tracker path' from a wide range of initial conditions.

\item
An interesting alternative form of dark energy is provided by the
{\bf Chaplygin gas} \cite{chap1,chap2}
which obeys the equation of state
\beq
p_c = - A/\rho_c~.
\label{eq:chap_state}
\eeq
The energy density of the Chaplygin gas evolves according to
\beq
\rho_c=\sqrt{ A+B (1+z)^6}\,\,,
\eeq
from where we see that $\rho_c \to \sqrt A$ as $z \to -1$ and $\rho_c
\to \sqrt{B}(1+z)^3$ as $z \gg 1$. Thus, the Chaplygin gas behaves
like pressureless dust at early times and like a cosmological constant
during very late times.

The Hubble parameter for a universe containing cold dark matter and
the Chaplygin gas is given by
\beq\label{eq:hub_chap}
H(z) = H_0\left\lbrack \om(1+z)^3 + \frac{\om}{\kappa}\sqrt{\frac{A}{B} +
(1+z)^6}\right\rbrack^{1/2}\,\,,
\eeq
where $\kappa = \rho_{0m}/\sqrt{B}$.
It is easy to see from (\ref{eq:hub_chap}) that
\beq\label{eq:chap_def}
\kappa = \frac{\rho_{0m}}{\rho_c}(z \to \infty)\,\,.
\eeq
Thus, $\kappa$ defines the ratio between CDM and the Chaplygin gas
energy densities at the commencement of the matter-dominated stage.

\end{itemize}
For all of the above models the
statefinders $r$ and $s$ can be easily expressed in terms of the
Hubble parameter $H(z)$ and its derivatives as follows:
\ber\label{eq:statefinder}
r(x) &=& 1-2 \frac{H^{\prime}}{H} x+\left\lbrace\frac{H^{\prime\prime}}{H}
+\left(\frac{H^{\prime}}{H}\right)^2\right\rbrace x^2,\nonumber\\
s(x) &=& \frac{r(x)-1}{3 ( q(x)-1/2)}\,\,,
\eer
where $x=1+z$ and $H$ is given by (\ref{eq:quiessence}),
(\ref{eq:kinessence}), (\ref{eq:hub_chap}), (\ref{eq:hubble_brane})
for the different dark energy models.
The results, shown in figures \ref{fig:evolve} \& \ref{fig:brane}, demonstrate
that the Statefinder diagnostic can successfully differentiate between
dark energy models as diverse as LCDM,
Quintessence, the Chaplgyn Gas and Braneworlds (see \cite{statefinder} for other
recent applications of
the Statefinder diagnostic).

\section{Bouncing Braneworlds}
\label{sec:bounce}

A remarkable feature of Braneworld cosmology is that the expansion dynamics of
the early universe can be {\em non-singular}. Indeed, as shown in \cite{ss03}
bouncing Braneworld models can be constructed, based upon a Randall--Sundrum
type action, but in which the extra dimension is {\em time-like}. In this case
one starts from the action \cite{ss03}
\begin{equation} \label{eq:bounce_action}
S = M^3  \left[ \int_{\rm bulk} \left({\cal R} - 2 \Lambda \right) - 2 \epsilon
\int_{\rm brane} K  \right] - \int_{\rm brane} 2 \sigma + \int_{\rm brane} L
(h_{ab}, \phi) \, .
\end{equation}
where the parameter $\epsilon = 1$ if the
signature of the bulk space is Lorentzian, so that the extra dimension is
spacelike, and $\epsilon = -1$ if the signature is $(-,-,+,+,+)$, so that the
extra dimension is timelike. The evolution equations resulting from (\ref{eq:bounce_action})
have the form \cite{ss03}
\begin{equation}\label{cosmo1}
H^2 + {\kappa \over a^2} = {\Lambda_{\rm eff} \over 3} + {8 \pi G_{\rm N} \rho
\over 3} + {\epsilon \rho^2 \over 9 M^6} + {C \over a^4} \, .
\end{equation}
Where $G_{\rm N} = \epsilon \sigma / 12 \pi M^6$ is the effective gravitational
constant, and $\Lambda_{\rm eff} = \Lambda / 2 + \epsilon \sigma^2 / 3 M^6$ is
the effective cosmological constant on the brane. For $\epsilon = 1$,
Eq.~(\ref{cosmo1}) reduces to the well-known Randall--Sundrum form (we ignore
$\kappa / a^2, \Lambda_{\rm eff}$ and $C / a^4$ which are unimportant for our
purposes)
\beq
H^2 = \frac{8 \pi G_{\rm N}}{3}\rho \left(1+\frac{\rho}{2\sigma}\right)~.
\label{eq:no_bounce}
\eeq
On the other hand if $\epsilon = -1$ then we obtain
\beq
H^2 = \frac{8 \pi G_{\rm N}}{3}\rho \left(1-\frac{\rho}{2|\sigma|}\right)~,
\label{eq:bounce}
\eeq
where the brane tension $\sigma < 0$.
(The two braneworld models (\ref{eq:no_bounce}) and (\ref{eq:bounce}) are dual,
as demonstrated in \cite{copeland04}.)
Equation (\ref{eq:bounce}) clearly demonstrates that $H \simeq 0$ when
$\rho_{\rm bounce} = 2|\sigma|$
\ie the universe {\em bounces} when the energy density of matter becomes
sufficiently large.
It is important to note that the singularity-free nature of the early universe
is a generic outcome of this theory and does not depend upon whether or
not matter violates the energy conditions.
An example of a bouncing braneworld universe is shown in figure (\ref{fig:bounce}) below.

\begin{figure}[ht]
\centering
\resizebox{3.5in}{!}{\rotatebox{-90}{
\includegraphics[width=9cm]{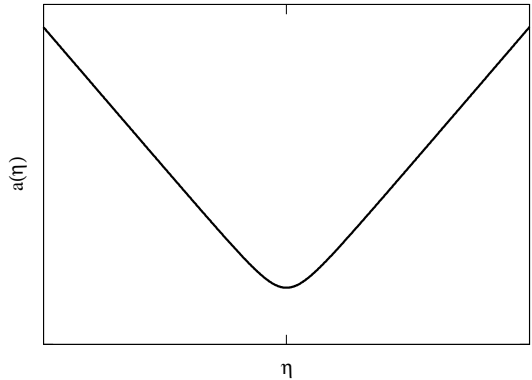}
}}
\caption{\small A bouncing radiation-dominated braneworld ($\eta = \int
dt/a$ is the conformal time). Figure courtesy of  \cite{ss03}.}
\label{fig:bounce}
\end{figure}

An early time bounce, such as (\ref{eq:bounce}), can be used to construct cyclic
models of the universe \cite{bfk04,kanekar,piao04a,piao04b}.
Consider for instance the mechanism proposed in \cite{bfk04}
in which Phantom dark energy is presumed to exist
in addition to matter and radiation so that
\beq
\rho = \rho_r + \rho_m + \rho_P
\eeq
where $\rho_r \propto a^{-4}$, $\rho_m \propto a^{-3}$,
$\rho_P \propto a^{-3(1+w)}$ and $w< -1$ in the case of Phantom leading to
\beq
\rho = \frac{A}{a^4} + \frac{B}{a^3} + Ca^{3(|1+w|)}~.
\label{eq:density_phantom}
\eeq
It is clear from (\ref{eq:density_phantom}) that $\rho$ grows at small
{\em as well as} large values of $a(t)$. Indeed
within one cycle the value of $H$ will pass through zero twice: (i) at early times
when the bounce in (\ref{eq:bounce}) is brought about due to large values of
the radiation density and (ii) during late times, when
the large value of the Phantom energy leads to $H=0$ in (\ref{eq:bounce})
and initiates the universe's recollapse.

\begin{figure}[ht]
\centering
\resizebox{4.5in}{!}{\rotatebox{-90}{
\includegraphics{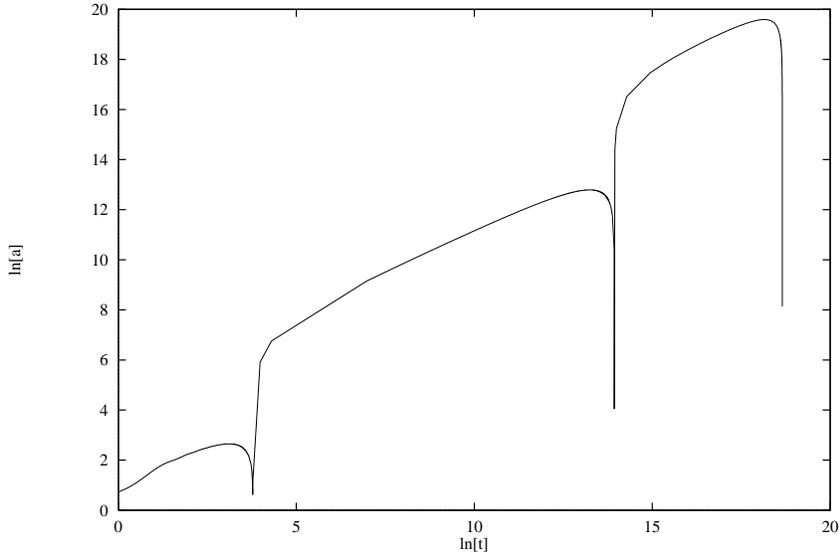}
}}
\caption{\small A cyclic universe with ever-increasing successive expansion
maxima. Figure courtesy of  \cite{kanekar}.} \label{fig:cyclic}
\end{figure}

Similarly, it is easy to show that a bouncing spatially closed universe with
matter satisfying $\rho + 3p > 0$ will also be `cyclic' in the sense that it
will pass through an infinite number of nonsingular expanding-collapsing epochs
\cite{kanekar,piao04a}. As demonstrated in \cite{kanekar,piao04a}, a massive
scalar field in such a universe usually leads to an increase in the amplitude
of consecutive expansion cycles and to a gradual amelioration of the flatness
problem; see figure \ref{fig:cyclic}.

\section{Conclusions}

Braneworld models hold interesting consequences both for the early as well as
the late-time evolution of the universe. At early times, the {\em increased
rate of expansion} in RSII type braneworlds allows scalar fields with steep
potentials to play the dual role of being the inflaton as well as Quintessence.

Braneworld models in which the RSII action is supplemented by an induced
gravity term on the brane  have important consequences for the {\em late-time}
evolution of the universe. For instance: (i)~ The effective equation of state
of dark energy in this scenario can be $w \leq -1$ as well as $w \geq -1$.
(ii)~The current acceleration of the universe can be a {\em transient\/}
phenomenon. (iii)~The braneworld universe can end its existence in a {\em
Quiescent\/} singularity at which the density, pressure and Hubble parameter
{\em remain finite}, while the deceleration parameter and all invariants of the
Riemann tensor diverge. (iv)~Braneworld models of dark energy can {\em
loiter\/} at high redshifts: $6 \lleq z \lleq 40$. The Hubble parameter {\em
decreases\/} during the loitering epoch relative to its value in LCDM. As a
result the age of the universe at loitering dramatically increases and this is
expected to boost the formation of high redshift gravitationally bound systems
such as $10^9 M_\odot$ black holes at $z \sim 6$ and lower-mass black holes
and/or Population III stars at $z > 10$, whose existence could be problematic
within the LCDM scenario. (v)~In addition to the above, an RSII type Braneworld
in which the extra dimension is time-like (instead of space-like) avoids the
initial Big Bang singularity by bouncing at early times. This property (which
holds even if matter satisfies all the energy conditions) has been used to
construct cyclic models of a singularity free universe.

\section{Acknowldgements}

It is a pleasure to thank Yuri Shtanov who collaborated with me on most of the
projects discussed in this paper.

\end{document}